# The leveled approach

## Using and evaluating text mining tools AVResearcherXL and Texcavator for historical research on public perceptions of drugs

Berrie van der Molen (UU), Lars Buitinck (NLeSC), and Toine Pieters (UU)

Utrecht University, Netherlands eScience Center

In our research on public perceptions of drugs in Dutch newspapers we have developed a leveled explorative historical research approach. We employ digital tools as "signposts" that indicate existing debates in newspapers that can be interpreted historically using a hermeneutic approach. Conceptualizing the ways we use text-mining tools as historians helps to align user needs with technological options.

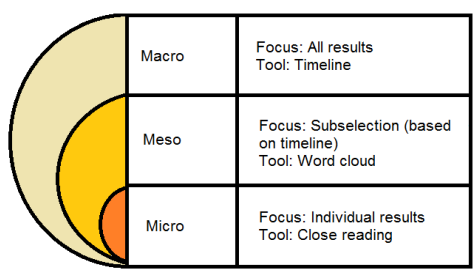

**Fig. 1.** Schematic representation of the leveled approach

The approach is based on continuous navigation between three levels of analysis (see Figure 1). All levels exist in AVResearcherXL and Texcavator, two digital text-mining tools that search Optical Character Recognition data of the approximately 11 million digitized newspaper pages from the Royal Dutch Library's archive. On each of the three levels, we use a specific visualization tool or analysis technique. The combination of text mining and visualization can lead to new historical insights (Van Eijnatten et al. 2014, 64). Here we show how our approach works using the example of public debates on amphetamine in The Netherlands between 1945 and 1969.

First, we develop a search query that encompasses known spelling variations and (brand) names of amphetamines:

```
am*etami* OR wekami* OR benzedri* OR perviti* OR perveti* OR
me*ylam*etami* OR neo-pharmedri* OR isophan OR neopharmedri
```

By incorporating a large number of spelling variations and (brand) names of amphetamine, we attempt to find all newspaper articles about the drug. On the *macro* level, we search the overall specified period (1945-1969) and we use the *timeline* functionality based on word frequency patterns. The pattern's peaks and valleys on the time-line help us to demarcate of sub-periods. The query is then entered again for each of the shorter sub-periods. The results for each of these sub-periods form the *meso* level of analysis. We compare the *word clouds* generated for each of the isolated periods, looking for strongly associated terms that indicate the existence of possible debates. On the *micro* level, we *read* and analyze the actual results (articles) yielded by the query altered with a significant term for a sub-period.

The analysis levels move between distant reading (Moretti 2013) on the macro and meso levels and close reading on the micro level. Distant reading helps to find relevant material for close reading. Gerben Zaagsma asserts that 'digital history' should not be seen as independent from 'non-digital history': as historians we should be hybrid (2013). But the navigation between distant and close reading and the relevant functionalities of digital tools can affect historical research results (Walma 2015, 67). Therefore we aim to be transparent about this process.

The strategies employed on each of the levels enable us to make an informed selection. Navigation between the levels can be based on temporal changes (zooming in on a broad query or zooming out to look at a more narrow query) or topical changes (adding related search terms to trace debates). The tools are not employed to answer questions directly; instead they help in formulating questions and guiding our focus. Distant reading functions as an incentive for qualitative analysis. Van Gorp et al. described AVResearcherXL's potential to 'deconstruct the archive, to dig deeper into trends, to discover new objects, and [...] to raise additional research questions' (51) – our research approach maximizes this potential.

Preliminary results from the amphetamine query indicate that initial amphetamine debates were mostly medical. This is not surprising, since the drug was freely available at pharmacies; the Dutch only started regulating amphetamine in 1969 (De Kort 1995, 175). Debates of abuse and addiction only seem to appear in the late 1960's. The meso level results thus indicate a shift in our overall (macro) period. Adding terms to the query enables us to confirm whether terms associated with abuse, for instance, did not occur much before the late 1960's. There is also reason to expand the time period on the macro level to see if the start of regulation led to more amphetamine addiction debates. These are examples of new pathways and alterations of our scope based on navigation between the analysis levels, leading to informed new selections of results.

The text-mining tools can be compared effectively by evaluating two usability factors inherent to our approach. The first usability factor is navigability. The benefits of the digital tools are most evident when they enable easy navigation between the three levels. The ability to switch between the overall time frame and smaller time frames and the ability to alter the queries with new terms (all based on results gained from using the digital tools) are essential in maximizing the signpost function of the digital tools. Examples that were helpful in this regard are AVResearcherXL's conjunctive query function (one click to start a new query with an additional term found in the word clouds) and Texcavator's clickable bars with relevant word clouds (creating successive meso levels with one click).

The second usability factor is search history. Search sessions are significantly more effective with the option to browse previous searches (which is accommodat-ed explicitly in Texcavator). Moreover, access to previous searches for different periods can help to make navigation between the levels of analysis easier. The stability of search settings is important as well. In AVResearcherXL we lost time because we had to re-set the time sliders repeatedly; in Texcavator we lost time because we had to re-set all filters when we wanted to add a search term to our query.

Explicit conceptualization thus helps us to compare the tools. Communication between users and developers of these tools benefits from conceptualization too. We emphasize the important role of domain users in the development of text- and data- mining technology and the importance of articulating and aligning the needs of the users with the technological options. This iterative process aimed at maximizing the big data potential requires continuous cooperation between historians and computer scientists and an open attitude regarding technological options and choices.